\RequirePackage{rotating}
\documentclass[sigconf]{acmart} 
\usepackage{amsmath}
\usepackage[ruled,vlined,linesnumbered]{algorithm2e}
\usepackage{amsmath}
\usepackage[T1]{fontenc}
\usepackage{graphicx}
\usepackage{textcomp}
\usepackage{tikz}
\usepackage{hyperref}
\usepackage{textcomp}
\usepackage{color}
\usepackage[utf8]{inputenc}
\usepackage{fancyvrb}
\usepackage{xcolor}
\usepackage{colortbl}
\usepackage{hhline}
\usepackage{lscape}
\usepackage{upquote}
\usepackage{array}
\usepackage{adjustbox}
\usepackage{rotating}
\usepackage{stmaryrd}
\usepackage{pifont}
\usepackage{multirow}
\usepackage{wrapfig}
\usepackage{verbatim}
\usepackage{colortbl}
\usepackage{rotating}
\usepackage{framed}
\usepackage[strict]{changepage}    

\SetKwProg{Proc}{Procedure}{}{end}
\SetKw{Continue}{continue}
\SetKw{Break}{break}

\newcolumntype{L}[1]{>{\raggedright\arraybackslash}m{#1}}
\newcolumntype{C}[1]{>{\centering\arraybackslash}m{#1}}
\newcolumntype{R}[1]{>{\raggedleft\arraybackslash}m{#1}}

\fvset{commandchars=\\\{\}}

\definecolor{Green}{rgb}{0.0, 0.56, 0.0}
\definecolor{Gray}{gray}{0.85}
\newcommand{\Blue}[1]{\textcolor{blue}{\textbf{#1}}}
\newcommand{\Green}[1]{\textcolor{Green}{\textbf{#1}}}

\newcommand{\equ}{$\equiv$}

\newcommand{\CodeIn}[1]{\begin{small}\texttt{#1}\end{small}}
\newcommand{\Comment}[1]{}
\newcommand{\NoType}[1]{} 
\newcommand{\Space}[1]{}

\newcommand{\DefMacro}[2]{\expandafter\newcommand\csname rmk-#1\endcsname{#2}}
\newcommand{\UseMacro}[1]{\csname rmk-#1\endcsname}

\DefMacro{check}{\ding{51}}
\DefMacro{corr}{\ding{72}}
\DefMacro{plausible}{\ding{51}}

\newcommand\diff\setminus

\definecolor{formalshade}{rgb}{0.95,0.95,1}
\definecolor{mygray}{rgb}{0.80,0.80,1}
\definecolor{cadetgrey}{rgb}{0.57, 0.64, 0.69}
\definecolor{silver}{rgb}{0.75, 0.75, 0.75}
\definecolor{whitesmoke}{rgb}{0.96, 0.96, 0.96}

\usepackage{graphicx}
\begin{document}

\title{Structure Editor for Building Software Models}

\author{Mohammad Nurullah Patwary}
\affiliation{%
  \institution{The University of Texas at Arlington}
  \country{Arlington, TX, USA}
}
\email{mxp9161@mavs.uta.edu}

\author{Ana Jovanovic}
\affiliation{%
  \institution{The University of Texas at Arlington}
  \country{Arlington, TX, USA}
}
\email{ana.jovanovic@mavs.uta.edu}

\author{Allison Sullivan}
\affiliation{%
  \institution{The University of Texas at Arlington}
  \country{Arlington, TX, USA}
}
\email{allison.sullivan@uta.edu}



%
%


%
%

%
%

\begin{abstract}
Alloy is well known a declarative modeling language. A key strength of Alloy is its scenario finding toolset, the Analyzer, which allows users to explore all valid scenarios that adhere to the model’s constraints up to a user-provided scope. Despite the Analyzer, Alloy is still difficult for novice users to learn and use. A recent empirical study of over 93,000 new user models reveals that users have trouble from the very start: nearly a third of the models novices write fail to compile. We believe that the issue is that Alloy's grammar and type information is passively relayed to the user despite this information outlining a narrow path for how to compose valid formulas. In this paper, we outline a proof-of-concept for a structure editor for Alloy in which user's build their models using block based inputs, rather than free typing, which by design prevents compilation errors. 
\end{abstract}

\maketitle


\section{Introduction}\label{sec:intro}

Despite their ability to help developers produce correct software, software models are not widely adopted. Unfortunately, modeling languages are notoriously difficult to learn, which is compounded by development toolsets that lag behind the state-of-the-art for integrated development environments (IDE)~\cite{fmedu_tools,humanin_fm}. 
Alloy is a well known modeling language that has been used to help validate software designs~\cite{CD2Alloy,Margrave,ChordAlloy,WickersonETAL2017,ChongETAL2018}, to test and debug code~\cite{DiniETALKoratAPI2018,MarinovKhurshid01TestEra,SamimiETALECOOP2010,ZaeemKhurshidECOOP2010} and to provide security analysis of systems~\cite{CheckMateMicro2019,websecurity,BagheriETAL2018}. Alloy has a reputation for being a more user friendly formal method, mainly due to its IDE the Analyzer, which lets users explore their models by producing a collection of satisfying instances that visually display allowed and prevented behavior. Over the years, there have been several efforts to build introductory formal methods courses centered around Alloy~\cite{alloystudy1,alloystudy2,alloystudy3}. Unfortunately, these studies have revealed that even with the Analyzer, new users still feel overwhelmed trying to learn the language. 

One reason for this is that several studies, including those done outside of an educational setting~\cite{spincomparison,alloyformalise23study}, have revealed that some of the commonly sighted strengths of the Analyzer are not actually  helpful to new users. While one perceived strength, visualized instances, has received a lot of attention~\cite{abstractalloy,Hawkeye,Reach,Amalgam,Aluminum,PorncharoenwaseETALCompoSat2018}, a recent empirical study exploring over 93,000 Alloy models written by novice users points to another ``strength'' that can be burdensome: the Analyzer's compiler-based error reports~\cite{jovanovic2024empirically}. The Analyzer uses Alloy's set of grammar rules as well as an internal type system to flag syntax and type based compilation errors. 
If an error is detected, then the user is given a summary message and problematic portions of the model's text are highlighted in red.

Unfortunately, the guidance provided by the compiler's error reports can be ambiguous at best, with the text report templates being too generic, and misleading at worst, with the highlighted formulas compiling in isolation. The lack of quality feedback provided by these error reports is a cascading problem, as this study reveals that users struggle with the basics of writing a model: 72.75\% of all users make at least one compilation mistake and 29.28\% of the models overall fail to compile. Furthermore, users can get stuck in long edit chains trying to make their model compile. 7.48\% of the time users need 5 or more edits to make their model compile, and users even give up without making a compilable model 2.88\% of the time. In order to help developers write valid models faster and easier, a new feedback mechanism is needed.


\textit{Our theory} is that Alloy's grammar and type rules can help users determine how to compose their formulas, as these rules combine together to provide an underlying rigid environment to build formulas in. However, presenting this information at compile time is too passive and muddles the transfer of knowledge. \textit{Our vision} is to enable the user to proactively see the grammatical and type implications of their choices by building models in a structure editor. Structure editors enable users to build programs not by free typing in an editor, but instead by interacting with a tree-structured representation, eliminating the possibility of introducing syntactic errors. Unfortunately, structure editors themselves are not widely adopted, as existing editors have a difficult time balancing the ease of making edit actions with guaranteeing the structural validity of the code~\cite{bahlke1992design,bau2017learnable,holwerda2018usability}. However, given that a user's options on what to write next to finish composing a formula are more restricted by Alloy's grammar and type system than other languages with structure editors, and that new users struggle with the grammar, we believe that a structure editor for Alloy will be beneficial. 

\usetikzlibrary{shapes.geometric, arrows}
\tikzstyle{arrow} = [line width=1.5pt,->,>=stealth]
\tikzstyle{darrow} = [line width=1.5pt,<->,>=stealth]
\tikzstyle{ListAtom} = [rectangle, minimum width=1.25cm, minimum height=.7cm, text centered, draw=black, fill=orange!75!yellow]
\tikzstyle{Trash} = [rectangle, minimum width=1.25cm, minimum height=.7cm, text centered, draw=black, fill=white!50!red]
\tikzstyle{Protected} = [rectangle, minimum width=1.25cm, minimum height=.7cm, text centered, draw=black, fill=yellow!75!orange]
\tikzstyle{State} = [circle, minimum width=0.5cm, minimum height=0.5cm, text centered, draw=black, fill=white]

\begin{figure}
\begin{center}

\scriptsize
\begin{Verbatim}[numbers=left,xleftmargin=6mm]
\Blue{var sig} File \{ \Blue{var} link : \Blue{lone} File \}
\Blue{var sig} Trash \Blue{in} File \{\}
\Blue{var sig} Protected \Blue{in} File \{\}

\Green{/* All unprotected files are deleted.*/}
\Blue{pred} inv5 \{ \Blue{all} x : File | x \Blue{not} \Blue{in} Protected => x \Blue{in} Trash \}
\Green{/*Chain of edits from one session from the Alloy4Fun dataset.*/}
inv5 S: \Blue{all} x : {\setlength{\fboxsep}{1pt}\colorbox{red!30}{!\Blue{in}}} Protected | x \Blue{in} Trash
inv5 S: \Blue{all} x : File | x !\Blue{in} Protected {\setlength{\fboxsep}{1pt}\colorbox{red!30}{|}} x \Blue{in} Trash
inv5 S: \Blue{all} x : File | {\setlength{\fboxsep}{1pt}\colorbox{red!30}{\Blue{all}}} x !\Blue{in} Protected | x \Blue{in} Trash
inv5 T: \Blue{all} x : File | x {\setlength{\fboxsep}{1pt}\colorbox{red!30}{!\Blue{in}}} Protected -> x \Blue{in} Trash

\Green{/* The protected status never changes.*/}
\Blue{pred} inv10 \{ \Blue{always} Protected = Protected' \}
\Green{/*Chain of edits from Alloy4Fun Dataset.*/}
inv10 T: {\setlength{\fboxsep}{1pt}\colorbox{red!30}{\Blue{always}}} Protected \Blue{once} Protected
inv10 T: {\setlength{\fboxsep}{1pt}\colorbox{red!30}{\Blue{after}}} Protected \Blue{always} Protected
inv10 T: {\setlength{\fboxsep}{1pt}\colorbox{red!30}{Protected}} \Blue{since} Protected
inv10 T: {\setlength{\fboxsep}{1pt}\colorbox{red!30}{\Blue{always}}} Protected \Blue{since} Protected
inv10 T: \Blue{always} ({\setlength{\fboxsep}{1pt}\colorbox{red!30}{Protected}} \Blue{since} Protected)
\end{Verbatim}

\end{center}

\caption{Alloy Model of a File System Trash Can}
\label{fig:trash}
\end{figure}


\section{Motivating Example}\label{sec:bg}

In this section, we outline key concepts of Alloy and the different types of compilation errors. 

\subsection{Model of a File System Trash Can in Alloy}

Figure~\ref{fig:trash} displays a model of a file system trash can from the Alloy4Fun dataset~\cite{alloy4funbenchmark}. Alloy4Fun is an online platform in which users fill in empty predicates to match a given English description. 
Signature paragraphs introduce named sets and can define relations, which outline relationships between elements of sets. Line 1 introduces a named set \CodeIn{File} and establishes that each \CodeIn{File} atom connects to zero or one (\CodeIn{lone}) \CodeIn{File} atoms through the \CodeIn{link} relation. Lines 2 and 3 introduce the named sets \CodeIn{Trash} and \CodeIn{Protected} as subsets (\CodeIn{in}) of \CodeIn{File}. All signatures and relations are  mutable (\CodeIn{var}), which means that the elements of these sets can vary across different states. 

Predicates (\CodeIn{pred}) introduce named formulas that can be invoked elsewhere. Predicate \CodeIn{inv5} uses universal quantification (\CodeIn{all}), subset (\CodeIn{in}), subset exclusion (\CodeIn{!in}) and implication (\CodeIn{=>}) to express that if a file is not in the set of protected files, then the file is in the trash. Predicate \CodeIn{inv10} uses the future temporal operator (\CodeIn{always}) and set equality (\CodeIn{=}) to express that the set of protected files is always the same. To execute predicates, users write Alloy commands. Commands have a scope that places an upper bound on the size of all signature sets. The command ``\CodeIn{run inv5}'' searches for an assignment to all sets in the model using up to 3 File atoms by default.

\subsection{Non-compilable Alloy4Fun Submissions}

To illustrate what compilation errors look like, Figure~\ref{fig:trash} includes two chains of edits that represent back-to-back submissions made by a user through Alloy4Fun. `S' indicates a syntax error and `T' a type error. The red highlights mimics the highlights provided by the error report, which also includes a written message. 

\subsubsection{Syntax Errors}

The chain of edits for \CodeIn{inv5} depict a user trying to properly structure a quantified formula (lines 8-11). First, the user forgets to create the domain (line 8). For syntactic errors, the text portion of the error report will list expected tokens; however, this list can be too long to be effective. For line 8, the user is told there are 37 possible tokens expected in place of the highlighted text. After adding a domain, the error report highlights the misapplication of the vertical bar keyword on line 9, and informs the users that there are 38 possible tokens that could apply here. Yet, it takes two edits for the user to actually remove this incorrect syntax (lines 10-11). On line 11, the user replaces it with the cross product operator (\CodeIn{->}) when the user likely intended to use implication (\CodeIn{=>}). These two operators are visually similar, but logically very different. The error report does not direct the user to this mistake, but instead highlights the subset exclusion operator, as the set ``\CodeIn{x}'' and the set formed by ``\CodeIn{Product -> x}'' are not of the same type. 


\subsubsection{Type Errors}

The chain of edits for \CodeIn{inv10} depicts a user repeatedly making the same high level type error. Type based error reports provide varying degrees of feedback. For line 11, the report gives the detailed context that ``\CodeIn{Left type = {this/File}. Right type = {this/File->this/File}}'' in the error report.  Unfortunately, for lines 16-20, the report is minimal and only states that ``\CodeIn{this expression failed to typecheck}.'' Fundamentally, the underlying mistake is that \CodeIn{Protected} evaluates to a set, which is not inherently something that is true or false. Meanwhile, the preceding temporal operators expect to reason over a boolean formula. To illustrate, consider evaluating the following formulas if \CodeIn{Protected = File0}:

\begin{footnotesize}
\begin{Verbatim}[frame=lines,rulecolor=\color{lightgray}]
\Blue{always} (Protected) \equ \Blue{always} (\{File0\}) \equ type error
\Blue{always} (\Blue{some} Protected) \equ \Blue{always} (\Blue{some} \{File0\}) \equ true
\end{Verbatim}
\end{footnotesize}

\noindent Since this hanging use of \CodeIn{Protected} is repeated across all the edits, the error reports never help the user realize the mistake. Case in point, on line 20, the user resorts to trying to insert parenthesis to resolve the issue before giving up.

\begin{figure}
    \centering
    \includegraphics[scale=0.45]{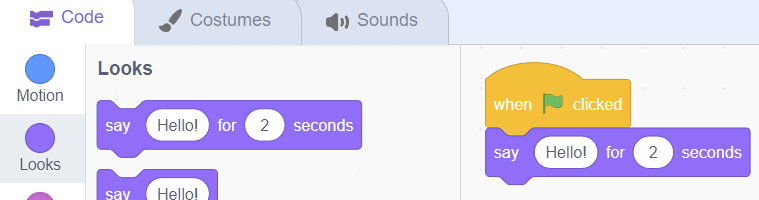}
    \caption{Hello World in Scratch}
    \label{fig:scratch}
\end{figure}

\section{Design Overview}\label{sec:tech}

In this section, we outline a proof of concept for a structure editor for Alloy. Figures~\ref{fig:inv5se} and \ref{fig:inv10se} display snapshots of using a structure edit to construct \CodeIn{inv5} and \CodeIn{inv10} respectfully, which we will use throughout this section to highlight how our design can prevent compilation errors and even guide development. 

\subsection{Blocked Based Editor}

Block based editors, which are currently the most popular type of structure editor, have users build programs by using a drag-n-drop interface to connect blocks of code. Each block typically represents a syntactic form within the language, e.g. an if statement. The visual design of blocks can provide valuable context. For instance, the shape of blocks can indicate how blocks interconnect. Figure~\ref{fig:scratch} displays a hello world program in Scratch, one of the most popular block based editor~\cite{maloney2010scratch}. ``Event'' code blocks are yellow and represent the start of a program, given by its shape which allows blocks to be added to the bottom but not the top. The interlocked purple block is a ``Looks'' code block that has a character say ``hello.''  

\begin{figure}
\begin{center}

   \includegraphics[width=7.3cm]{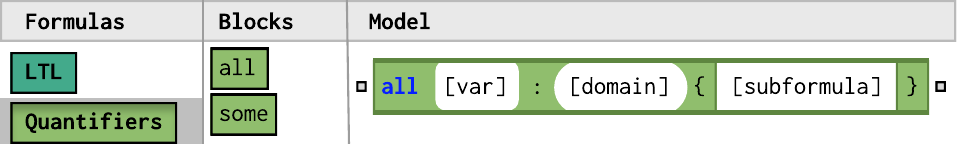} 
   
     \footnotesize\textbf{(a) - prevents Figure~\ref{fig:trash} line 8 error} 
     
     \includegraphics[width=7.3cm]{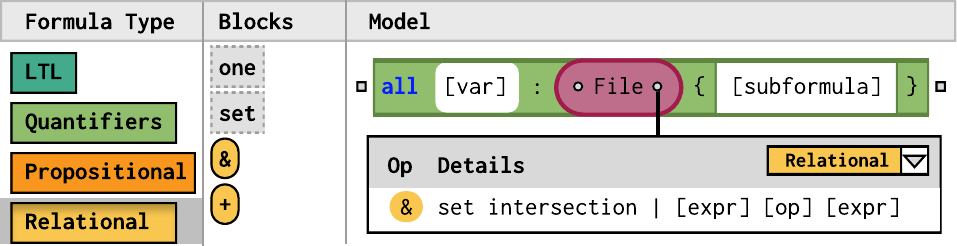} 
     
     \footnotesize\textbf{(b) - guides correction of Figure~\ref{fig:trash} line 8 error} 
     
     \includegraphics[width=7.3cm]{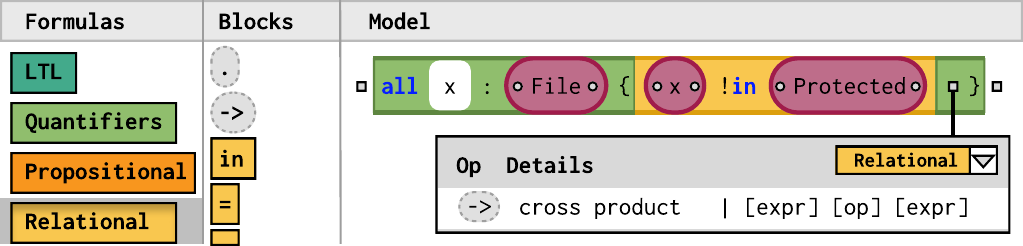} 
     
     \footnotesize\textbf{(c) - prevents Figure~\ref{fig:trash} line 11 error} 
     
     \includegraphics[width=7.3cm]{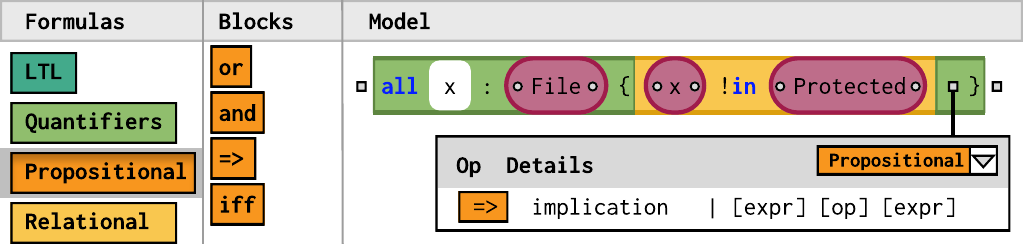} 
     
     \footnotesize\textbf{(d) - guides correction of Figure~\ref{fig:trash} line 11 error} 

\end{center}

\caption{Wireframe of structure editor steps for inv5}

\label{fig:inv5se}
\end{figure}

Scratch has been successfully used in introductory programming courses, even at the K-12 level~\cite{su2022effect,duo2023analysis,montiel2021educational}. This later point is of interest to us, as one of the main applications we envision is to help users learn Alloy faster and easier than is currently possible using the Analyzer. Therefore, our structure editor is inspired by the best practices derived from Scratch's success. As figures~\ref{fig:inv5se} and~\ref{fig:inv10se} show, like Scratch, our structure editor is composed of three interconnected panels. The ``Formula Type'' panel displays different categories of blocks and the ``Blocks'' panel displays all selectable blocks of that type, which can be dragged and dropped into the ``Model'' panel canvas to build a formula. 

The use of blocks themselves is already helpful at preventing errors, as blocks remove typos, which would prevent the type error on line 11 in Figure~\ref{fig:trash}. The difference between the implication and the cross product operators is two similar looking characters that are also neighboring keys on a keyboard, increasing the chance a user might mix them up. However, in a structure editor the user would instead select the implication operator and not type it, as Figure~\ref{fig:inv5se} (d) illustrates. While this mistake throws an error, it is possible for the user to make a typo that compiles, such as mixing up set intersection and set difference, which are also neighboring keys. Moreover, if the user actually intended to use the cross product operator, our structure editor will prevent this placement, since as Figure~\ref{fig:inv5se} (c) illustrates, the operator would not be selectable.

\subsection{Definition of Blocks}
We uses operators and basic sets as the selectable blocks for construction. For operators, Alloy's underlying grammar divides operators based on structure only, e.g. unary or binary. However, we believe structure alone is too broad of a category for user to easily find the desired operator. For instance, the binary structure would intermix linear temporal operators (e.g. \CodeIn{since}) with relational operators (e.g. \CodeIn{+}) and with propositional operators operators (e.g. \CodeIn{or}). Instead, we plan to organize operators by type of mathematical logic, as outlined in Figure~\ref{fig:operators}, then by structure. This organization may change depending on future user studies. For basic sets, the available basic sets for a hole are all signatures, all relations, and any locally defined variables within scope of the formula edit. 

Scratch's puzzle piece block shape design guides users to build sequential steps, which is appropriate for an imperative language.
However, Alloy is declarative, which means there is no control flow. Therefore, our shape design does not convey order of execution, but we do use our block shape to convey whether a formula produces a set (rounded) or boolean (squared). The shapes are used throughout the ``Blocks'' and ``Model'' panels, helping to convey the underlying type structure for construction as well as already built formulas. 

\begin{figure}
\centering
\begin{footnotesize}
\begin{Verbatim}[frame=lines,rulecolor=\color{lightgray}]
\textbf{Relational Logic Operators:}
unOp::= ~ | * | ^ | ! | no | lone | some | one | set 
binOp::= & | + | - | ++ | <: | :> | . | -> | in | = | < | > | =< | >=
\textbf{Propositional Logic Operators:}
binOp::=  or | and | iff | => 
\textbf{First-order Logic Operators:}
quant::= all | no | some | lone | one
\textbf{Linear Temporal Logic Operators:}
unOp::= always | eventually | after | before | historically | once | '
binOp::=  since | triggered | until | releases | ;
\end{Verbatim}
\end{footnotesize}
    \caption{Division of Operators for Structure Editor}
    \label{fig:operators}
\end{figure}


\subsection{Block Based Construction}\label{con}

To construct a formula, users add blocks at open locations, which we will refer to as holes, within the model. Holes are indicated by either (1) a labeled bracket, e.g. \CodeIn{[domain]} in Figure~\ref{fig:inv5se}~(a), or (2) circle or square icon, e.g. $\circ$/\footnotesize{$\square$}\normalsize, that represent extension points on existing formulas. 
A model starts off as a single square extension point. To add blocks, we plan to explore using (1) the traditional drag-and-drop interface and (2) a pop-up menu, as seen in Figure~\ref{fig:inv5se}~(b), that allows users to select a block at a specific hole. As Section~\ref{syntax} will detail, not all type errors can be prevented by distinguishing sets from booleans and we would need an intractable number of shapes to account for all possible types that could be introduced. Therefore, we plan to have the user select an active hole. This enables us to efficiently make blocks that throw errors unselectable, which we indicate as grayed out blocks, as is the reason we plan to explore a popup menu interface. We could prevent the user from placing a block into a hole after selection, but we believe a proactive indicator will help users more easily determine the right block to match their intent and help educate the user on how to compose formulas. 

\subsubsection{Preventing Syntax Errors} The design of our two holes ensure users build models that are syntactically valid. The labeled bracket enables us to let the user know the structural expectations for a formula. These labeled brackets are similar to Scratch's holes, as seen in Figure~\ref{fig:scratch} where the Look block lets the user add ``hello'' and ``2'' after placement. Figure~\ref{fig:inv5se}~(a)'s labeled brackets conveys to the user that they will need to define a variable, a domain and a subformula in order to create a universally quantified formula. This directly prevents the syntax error from line 8 in Figure~\ref{fig:trash} from being possible. The extension points enable us to convey to the user all the locations in which the current formula can be extended. For instance, Figure~\ref{fig:inv5se}~(b) shows how the user can select the latter extension point for \CodeIn{File} and build out a more intricate domain, such adding a set intersection block to make ``\CodeIn{File \& Trash}.''  Furthermore, these extension points do not force the user to build the formula in a specific order. For instance, in Figure~\ref{fig:inv5se}~(b), the user can select the leading extension point and build the domain ``\CodeIn{Trash \& File}.'' 

\begin{figure*}
\begin{center}
\begin{tabular}{c|c}
   \includegraphics[width=7.3cm]{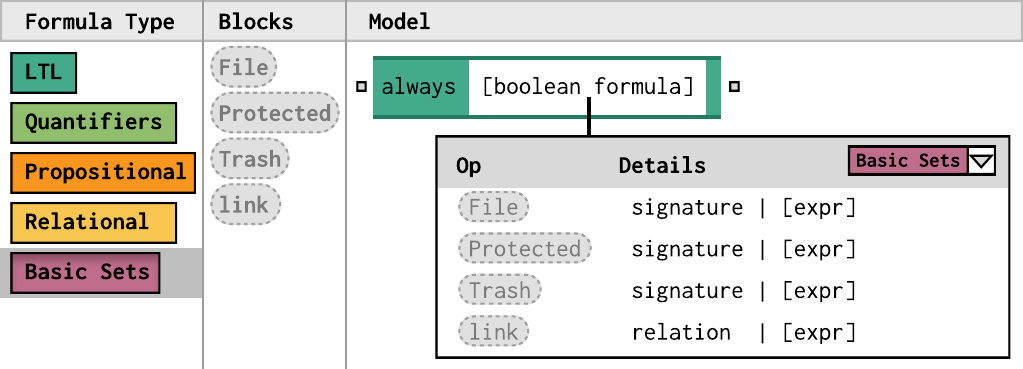}  & \includegraphics[width=7.3cm]{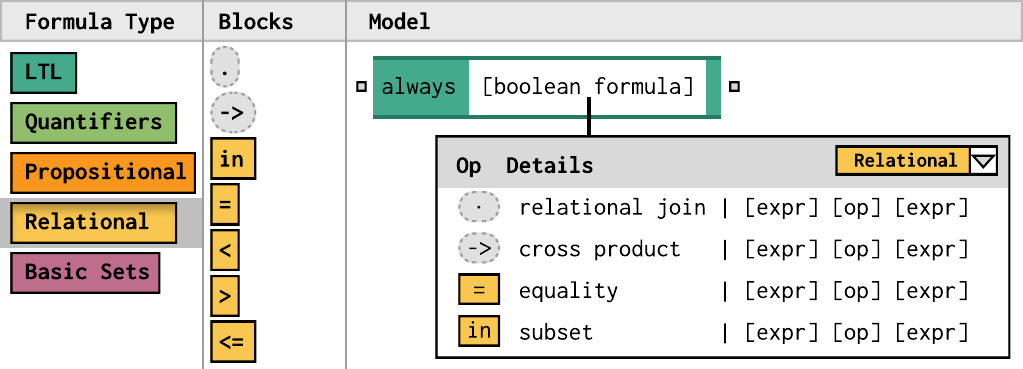} \\
     \footnotesize\textbf{(a) - prevents Figure~\ref{fig:trash} lines 16-20 error} & \footnotesize\textbf{(b) - guides correction of Figure~\ref{fig:trash} lines 16-20 error}
\end{tabular}
\end{center}
\vspace{-1ex}
\caption{Wireframe of structure editor selection options for building off of ``\CodeIn{always}'' in inv10}
\label{fig:inv10se}
\end{figure*}


\subsubsection{Preventing Type Errors}\label{syntax}

There are two classifications of type errors, primitive and violations of Alloy's type-checking rules. Primitive type errors occur when a user creates a formula that produces a set when a boolean is expected or vice versa. Our shape design handles these errors, as we will enforce that only rounded shapes can be inserted into rounded holes. We also propagate this information to the active hole substitution filtering, mentioned in Section~\ref{con}. To illustrate, when composing inv10, our structure editor prevents the user from directly adding any of the basic sets, as seen in  Figure~\ref{fig:inv10se} (a), as they produce sets when a boolean is needed. This immediately addresses the high level error the user was never able to correct in Figure~\ref{fig:trash} lines 16-20. Simply conveying the primitive type information  can have a big impact guiding users. To illustrate, Figure~\ref{fig:inv10se} (b) shows that the user will learn that set-based relational operators are unselectable but comparison-based relational operators are selectable. Likewise, Figure~\ref{fig:inv5se} (b) and (c) also show examples of this. 


However, shape alignment cannot prevent all type based errors. For instance, for the type error on line 11 in Figure~\ref{fig:trash}, ``\CodeIn{x}'' and ``\CodeIn{Protected->x}'' are both sets, but they have different internal types. 
Sets in Alloy have internal types related to their element's relationship with the signatures and relations of the model. 
Type-checking errors occur when sets of different internal types are used within the same subformula. We can use constraint checking to determine if inserting a given block into a hole will create a type conflict. This enable us to provide nuanced but helpful information to the user. For instance, consider the intermediate step ``\CodeIn{x !in [rhs]}'' to build the formula ``\CodeIn{x !in Protected}'' from Figure~\ref{fig:inv5se} (c). When listing the Basic Set blocks, we can prevent \CodeIn{link} from being selectable, as \CodeIn{x} is of type \CodeIn{File} and \CodeIn{link} is of type \CodeIn{File->File}. 
\section{Future Plans}

To bring our structure editor development environment from a proof of-concept to a reality, there are a few key problems to address.

\textbf{\textit{Feel of Workflow.}} 
First, edit actions to compose a model need to be intuitive, enabling the user to feel as if their construction of a model is not being unnecessarily restricted. One issue we forsee is that for construction, our extension points allow some flexibility, but this may not be enough. For instance, returning to the previous example, we could let the user first build ``\CodeIn{x !in File}'' in order to extend it into ``\CodeIn{x !in File.link},'' which may feel more natural than the user having to select the relational join first to create ``\CodeIn{File\textbf{.}link}.'' Therefore, we plan to explore interface designs which allow users to build intermediate formulas to then insert into holes. 
Besides construction, we will also need to enable more edit actions, including refactoring, replacements, and deletions.

For all edit actions, we will need to (1) ensure the edits happens without noticeable lag and (2) conduct user studies to ensure that the workflow to take these actions is not cumbersome. For the former, between the current performance of Alloy's compiler and our success at leveraging constraint checking for several Alloy tasks such as fault localization and repair~\cite{ARepair,AlloyFL,AUnit}, we believe responsiveness is feasible. For the latter, we anticipate using students as novice users and active members of the Alloy discourse group as expert users, and designing students to explore multiple design choices such as categorization and order of blocks, how blocks can be added and how users select different edit actions. 

\textbf{\textit{Adoption.}} 
Second, while a block-based editor is likely to have a beneficial impact for formal methods education, creating an editor that sees adoption even among expert users is an open problem. One avenue we plan to explore is to enable different edit modes, giving the user the option to choose how guided the structure editor will be. We current plan to investigate two options: (1) an active mode following the traditional structure editor workflow and (2) an overlay mode in which the user free types and the structure editor live compiles the user's formulas and flags issues as they appear. In the later, we will need to ensure we balance how often we attempt to compile to ensure that we flag issues in a timely manner without nagging the user as they are actively typing out an operator or set. Another avenue we plan to explore is utilizing Alloy itself to add context to edit options. For instance, if the user selects a refactoring edit action to swap ``\CodeIn{a in b}'' with ``\CodeIn{b in a},'' we can generate an instance that highlights the difference between the two. The ability to explore the impact of an edit before selecting the edit could motivate expert users to develop in our structure editor, as our editor could save time spot-checking formulas.
\section{Related Work}
Structure editors is an active research field~\cite{maloney2010scratch,voelter2012language,teitelbaum1981cornell,mcnutt2023projectional,omar2017hazelnut,asenov2017envision,Fructure}, with the most popular type being block based, in which the user authors programs by dragging and dropping blocks together on a canvas. While block based editors have seen success in educational settings, they are not widely used due how clunky development can be~\cite{bahlke1992design,bau2017learnable,holwerda2018usability}. As a result, there is a whole body of work looking to enable other forms of inputs mechanism largely centered around keystrokes~\cite{kolling2010greenfoot,berger2016efficiency,cursorless,moon2022tylr,moon2023gradual}. These other iterations are avenues we will consider for inputs if our user studies reveal blocks are not ideal. However, given the rigid development options enforced by Alloy's type and grammar system, we believe that a block-based editor is a good starting point. 
\section{Conclusion}\label{sec:con}
A recent empirical study has revealed that users struggle from the get go to write models: new users have a hard time producing models that successfully compile. This paper presents the concept of a structure editor development environment for Alloy that replaces the passive transfer of knowledge that compilation error reports provide with an active guided development environment. 

%
%
%
\bibliographystyle{splncs04}
\bibliography{bib}

\end{document}